\documentclass[12pt,preprint]{aastex}
\begin{document}

\title{Inferring physical conditions in interstellar clouds of H$_2$  }

\author{Matthew K. Browning\altaffilmark{1}, Jason Tumlinson, and J. Michael Shull\altaffilmark{1}}

\affil{CASA, Department of
Astrophysical and Planetary Sciences, University of Colorado, Boulder, CO 80309}

\altaffiltext{1}{Also at JILA, University of Colorado and National
Institute of Standards and Technology}
\begin{abstract}
We have developed a code that models the formation, destruction, radiative
transfer, and vibrational/rotational excitation of H$_2$ in a detailed
fashion.  We discuss generally how such codes, together with FUSE
observations of H$_2$ in diffuse and translucent lines of sight, may be
used to infer various physical parameters.  We illustrate the effects of
changes in the major physical parameters (UV radiation field, gas density,
metallicity), and we point out the extent to which changes in one parameter
may be mirrored by changes in another.  We provide an analytic formula for
the molecular fraction, f$_{H2}$, as a function of cloud column density,
radiation fields, and grain formation rate of of H$_2$.  Some diffuse and
translucent lines of sight may be concatenations of multiple distinct
clouds viewed together.  Such situations can give rise to observables that
agree with the data, complicating the problem of uniquely
identifying one set of physical parameters with a line of sight.  Finally,
we illustrate the application of our code to an ensemble of data, such as
our FUSE survey of H$_2$ in the Large and Small Magellanic Clouds
(LMC/SMC), in order to constrain the elevated UV radiation field intensity
and reduced grain formation rate of H$_2$ in those low-metallicity
environments.

\end{abstract}

\section{Introduction}
\subsection{Overview}
No molecule in astrophysics is as ubiquitous or as far-reaching in its
effects as molecular hydrogen (H$_2$).  Found in nearly every physical
environment and in every temporal domain, it was the first neutral molecule to
form after the Big Bang, it is a major constituent of giant molecular
clouds, and it forms the bulk of the atmospheres of Jovian planets.  Of
particular relevance to our interests is the fact that H$_2$ is the most
abundant molecule in the interstellar medium (ISM); nearly every target
through the Galactic disk and halo observed with the Far Ultraviolet
Spectrographic Explorer (FUSE) since its launch in June 1999 has exhibited
signs of H$_2$ along the line of sight (Shull et al. 2000). Studying H$_2$
is a major scientific goal of FUSE; in addition to the diffuse cloud
program, nearing completion with $\sim$ 100 targets in the Galactic disk, active FUSE campaigns exist to study H$_2$ in the LMC/SMC
(Tumlinson et al. 2002) and in denser ``translucent'' clouds having a
visual extinction in the range of about $A_V = 1 \rightarrow 5$ (Snow et
al. 2000; Rachford et al. 2001). Thoroughly understanding such clouds of
H$_2$, which are the raw ingredients out of which giant molecular clouds
and, later, stars form could lead to better comprehension of that process
or the physical nature of the ISM in general.

Here, we discuss new computational models of interstellar clouds of H$_2$
and their application to FUSE data.  Information on the theory and
construction of these models, a summary of their application to large FUSE
datasets, and interpretations of the results form the bulk of this paper.

\subsection{Background material}

A brief reminder of the physics of the H$_2$ molecule is in
order. Molecular hydrogen has quantized electronic, vibrational, and
rotational degrees of freedom, giving rise to a commensurate set of energy
levels.  The ground electronic state ($X^{1} \Sigma_{g}^{+}$) is split into
a number of vibrational levels (labeled $v=0, 1,..., 14$),
which are in turn split into rotational levels ($J=0, 1, ...$). Higher
electronic states (the next two of singlet symmetry are labeled $B^{1}\Sigma_{u}^{+}$ and
$C^{1}\Pi_{u}$) are also split into vibrational and rotational levels, and
absorptions from the ground state into these electronic states are referred
to as the Lyman ($B^{1}\Sigma_{u}^{+} \leftarrow X^{1} \Sigma_{g}^{+}$) or
Werner ($C^{1}\Pi_{u} \leftarrow X^{1} \Sigma_{g}^{+}$) bands.  Molecular
hydrogen in its ground electronic state has no permanent electric dipole
moment, and hence no dipole-allowed vibrational or rotational transitions.
Consequently, cold H$_2$ is most readily detected via observation of its
ultraviolet absorption in these electronic bands (Shull \& Beckwith 1982).

The Lyman and Werner bands occur in the far-ultraviolet portion of the
spectrum, and so their observation is entirely the purview of space-based
missions.  Early efforts in this vein (see Snow 2000, and references
therein) included rocket-borne spectrographs (e.g., Carruthers 1967), the
\emph{Copernicus} mission, the Interstellar Medium Absorption Profile
Spectrograph, the ORFEUS spectrograph, and the Hopkins Ultraviolet
Telescope.  The data we consider here are from FUSE -- the most recent
mission capable of attacking this problem -- the details of which are
described in Moos et al. (2000) and Sahnow et al. (2000).  FUSE is a
multi-channel Rowland circle spectrograph, with coverage of the spectral
range from approximately 912 to 1187 \AA\, and a spectral resolution of
approximately $R=20,000$. It is FUSE's ability to measure accurate column
densities in rotational and vibrational levels that makes possible the
detailed analysis of diffuse and translucent lines of sight described here.

\section{Modeling H$_2$}
\subsection{Overview}
The modeling process itself is principally a game of balance.  In a steady
state, those processes that populate a level must balance those that
depopulate it. In the larger sense, there also exists a balance between
formation of the molecule on grains and its destruction, primarily via
absorption in the Lyman or Werner bands, followed by decay approximately
$11 \%$ of the time to the vibrational continuum of the ground electronic
state.  The primary population mechanisms are spontaneous radiative decay
from higher levels, cascades from upper levels following absorption in the
Lyman or Werner bands, direct formation into the level, and collisional
excitation. The primary depopulation schemes of rotational states $J \geq
2$ in the ground vibrational level, for the conditions prevalent in the
ISM, are radiative absorption out of a given level, spontaneous radiative
decay, and collisional de-excitation.  At low temperatures, only reactive
collisions with charged particles (H$^{+}$, H$_{3}^{+}$) couple odd-$J$
(ortho) and even-$J$ (para) H$_2$, which otherwise behave as separate
species.  Absorption in the Lyman and Werner lines varies with depth in the
cloud, as these lines become ``self-shielded''; this results in decreased
photo-destruction rates.  Finding the steady-state level populations at each
point in a cloud is then a matter of determining the magnitude of all these
effects, and then solving the system of equations for the resulting
$n(v,J)$, the physical density in each rotational, vibrational level.

The first attempts to model clouds in this fashion were carried out by
Black \& Dalgarno (1973). Many later efforts in the same vein (Spitzer
\& Zweibel 1974; Jura 1975a,
Jura 1975b; Black \& Dalgarno 1976, 1977; Shull 1978; van Dishoeck \& Black 1986,
hereafter vDB; Black \&
van Dishoeck 1987; Sternberg 1988; Sternberg \& Dalgarno 1989; Draine \& Bertoldi 1996) have followed basically this prescription, with some
variations in the processes taken into account, the rates 
used for collisional and radiative processes, and the approximations made.

We have developed a C++ code that performs the analysis described above.
Its treatment of the radiative transfer in the Lyman and Werner lines is
reasonably sophisticated.  Rather than relying on analytical or
``equivalent width'' approximations to the attenuation at each depth step,
we simulate the clouds as isothermal slabs with an ambient radiation field
incident on one or two sides. We carry out numerical integrations over full
Voigt line profiles to calculate the absorption out of each $(v,J)$ level
at each of typically 500 depth steps.  (In the case where clouds are
illuminated on two sides, an iterative procedure is used to solve for the
radiation field at each depth step.)  A major advantage of this technique
is that it allows us to account in an accurate and natural fashion for line
overlap, expected to be of special importance in the high-column-density
translucent systems.  Figure 1, taken from FUSE observations of five
different targets, shows how serious line overlap can be, as $N(H_{2})$
approaches 10$^{21}$ cm$^{-2}$.  This procedure
also provides us with a simulated absorption spectrum at every depth step
in the cloud, which provides a graphical check to the physical modeling.

Figure 2 shows a sample of simulated spectra at two different depths in a
cloud of physical density of hydrogen nuclei $n_{H} = 70$ cm$^{-3}$ and
temperature $T=98$ K.  The dashed line in this figure corresponds to a
depth in the cloud where the total H$_2$ column is $1.2 \times 10^{20}$
cm$^{-2}$; the solid line corresponds to $1.2 \times 10^{16}$ cm$^{-2}$.
The primary result of a model such as this is a set of physical densities
$n(v,J)$ at every depth in the cloud, which can be integrated through the
cloud to provide a column density $N(v,J)$, in turn comparable with FUSE
observations.  For completeness, we note that the rates for collisional and
radiative processes employed in our models differ (in some cases
appreciably) from those used in some previous models, owing to more recent
and sophisticated quantum-mechanical calculations of those rates. For
radiative rates, we use the work of Wolniewicz, Simbotin, \& Dalgarno
(1998) and Abgrall et al. (1994); collisional rates for H$_2$-H and
H$_2$-H$_2$ interactions are from Le Bourlot et al. (1999).  Proton-H$_2$
collision rates are from Dalgarno, Black, \& Weisheit (1973), and Gerlich
(1990); the former's rate for the $J=1 \rightarrow 0$ transition is, for
computational reasons, used for the large grids of models discussed below,
but this has negligible effect on the resulting $N(J)$.  We treat H$^{+}$ as
a species with fixed abundance, $n_{H+}/n_{H} = 10^{-4}$, and assume a
cosmic-ray ionizing frequency of $2 \times 10^{-17}$ s$^{-1}$.  We assume a Doppler line-broadening parameter $b=5$ km s$^{-1}$ in all the
models discussed here.

In Table 1, we compare a set of reference models, computed using our code,
to other models in the literature (vDB).  The models from the literature have fixed
$N(H_{2})$; to match roughly these total $N(H_{2})$, the comparison models using our code
were run with $r=1.33$ pc.  Models in this table have $n_{H}=250$
cm$^{-3}$, $I = 2 \times 10^{-8}$ photons cm$^{-2}$ s$^{-1}$ Hz$^{-1}$, and
$T=$ 20, 40, 60, 100 K, to match models C1, C4, C7, and C10 in vDB.  While we do not obtain precisely the same
$N(J)$ as vDB for a given set of physical conditions, the differences between
our results are $<$ 20 \% in most cases.  These discrepancies are probably attributable to our
different treatment of the radiative transfer, coupled with different
values for the collisional and radiative rates involved in these
calculations.  

We wish to keep the models relatively well-constrained by the data, in
order to examine what happens when basic physical parameters like
temperature, density, and UV radiation field are altered.  Therefore, we
have chosen to make a number of simplifications to the modeling process.
The remainder of this section is devoted to highlighting and discussing
some of these approximations.

\subsection{Isothermality}
Our models have constant physical temperature and total physical density.
This is clearly an idealization, and for larger clouds it is probably a
poor one, but we see no clearly superior and viable characterization of the
diffuse clouds we observe.  While polytropic equations of state have been
used to model interstellar clouds (e.g., vDB), and appear to be a
reasonable approximation to their structure, they are not appropriate for
our purposes here.  Modeling clouds as polytropes introduces additional
parameters that cannot be regarded as fixed by observations; it also
obscures one of our central goals, to determine what happens to
observational diagnostics of these clouds as basic physical parameters are
changed.  Also, some of the diffuse clouds discussed here may be so far
from a pressure-bounded equilibrium that polytropes would be inappropriate
as a description of their structure.  While such clouds are also unlikely
to be isothermal, nor even indeed truly amenable to the steady-state
modeling performed here, it is natural to model them first with as simple a
parameterization as possible.  Other schemes in which temperature and
density vary -- e.g. a ``constant pressure'' model where the product $n T$
is held constant -- might also be potentially appropriate descriptions of
these structures, but no strong constraints on the temperature and density
distributions exist.  Although an \emph{ad hoc} distribution might provide
a good match to some of the data, it is too ill-constrained to be of
serious use.

We have not attempted to solve for the thermal balance of the clouds
self-consistently, because it would be computationally intensive
and because some facets of the thermal balance problem are not  
well understood.  In particular, the photoelectric effect on dust grains
is probably a major
heating source for the gas (Wolfire et al. 1995; Boulanger et al. 2000), but the details of that
heating mechanism for a given physical environment are poorly constrained.
Leaving temperature as a free parameter also allows us to explore what will
happen to the H$_2$ in clouds where the gas is anomalously heated (by,
e.g., thermal shocks or MHD wave-damping).

\subsection{The Slab Geometry} We simulate the clouds as slabs with a radiation field
incident on one or two sides; this is also clearly an idealization.
Interstellar clouds are rarely slabs, but again it is not
entirely clear how one might improve on this picture without vastly
exceeding our ability to constrain it.  Representing the clouds as spheres
rather than slabs might more closely approximate their true nature,
but it is appreciably more difficult computationally. Such a
representation also seems to us unlikely to change
qualitatively the nature of our results here, so we have not undertaken
it.  The slab approximation means that we are probably underestimating the
radiation field at points interior to the cloud, since presumably the real
clouds are finite in directions perpendicular to the line of sight, and
therefore admit some radiation from those directions.  This
additional radiation input, however, should be small in most cases;
qualitatively, we expect the additional radiation to be comparable to that
coming from the ``far side'' of a two-sided slab. To estimate this
effect, we point out Table 2, which shows column densities, $N(J)$, for an arbitrary
model, calculated using both one and two-sided slab approximations. The
additional radiation input provided by the far side has the expected effect
of boosting the higher-$J$ levels slightly, but the effect is modest
(less than $\sim$ 0.3 dex in $N(J)$ for all models calculated under both schemes).

\subsection{Radiative Transfer Issues}
We assume that the line and continuum attenuations of the radiation field
are separable, and we incorporate dust into the continuum attenuation.  We
assume the dust model of Roberge et al. (1981), which is a
forward-scattering grain model.  This choice of model has previously been
shown to have relatively little effect on the H$_2$ population (vDB).  The
radiation field at each depth in the cloud is discretized on a large grid
of frequency points. The total attenuation at each depth is calculated
at all of those frequency points by considering the contribution to the
total absorption cross-section from every line in the Lyman and Werner
bands, together with the continuum attenuation.

The radiation fields that illuminate our model clouds are flat in photon
space between 912 and 1120 \AA , e.g. $I_{\nu} = 2 \times 10^{-8}$ photons cm$^{-2}$ s$^{-1}$
Hz$^{-1}$.  The actual Galactic far-UV spectrum varies somewhat across the
wavelengths relevant here.  However, we have chosen not to include this variation
as input to the models, because for some of the environments considered
here (e.g., the low-metallicity LMC and SMC, or Galactic clouds exposed to
high radiation field), a Galactic spectrum is probably inappropriate.  In
Table 3, we compare a representative model with a flat spectrum, $I_{\nu}=$ 2 $\times 10^{-8}$ photons cm$^{-2}$ s$^{-1}$
Hz$^{-1}$, against one with an
incident spectrum of UV radiation designed to approximate more closely UV
starlight (Draine 1978),
\begin{equation}
  \lambda u_{\lambda} = (4 \times 10^{-14} {\rm \ ergs \ cm}^{-3}) \ \chi \lambda_{3}^{-5} (31.016
  \lambda_{3}^{2} - 49.913 \lambda_{3} +19.897)
\end{equation}

\noindent with $\chi=1.71$ and where $\lambda_{3} =
\lambda/10^{3}$ \AA .  The level populations of models exposed to these
different UV radiation fields are not identical, but the differences are
small relative to those discussed in the analysis of data below.  Note that
while we refer here to a spectrum of $I=$ 1 $\times 10^{-8}$ photons
cm$^{-2}$ s$^{-1}$ Hz$^{-1}$ as a rough ``Galactic mean'' UV radiation
field, the column densities for clouds exposed to the Draine (1978)
radiation field are closest to those with a flat spectrum somewhere in the
range $I=$ (2-3) $\times 10^{-8}$ photons cm$^{-2}$ s$^{-1}$ Hz$^{-1}$.

\subsection{Formation Process}
In the conditions prevalent in the ISM, H$_2$ is believed to form primarily on the
surfaces of dust grains (Hollenbach, Werner, \& Salpeter 1971).  While a few
details of this process remain unclear
(see Herbst 2000), the basic scheme is that formation proceeds
when an H atom collides with a dust grain, is adsorbed and moves  across
the grain surface, encounters a previously adsorbed H atom on the grain,
recombines, and pops off the grain.  Much recent work has provided new
insights into the details of this formation process (e.g., Katz et
al. 1999), although these turn out to be largely irrelevant to our
results -- see below.  In principle, the rate at which formation on
grains occurs is a function of the grain size, abundance, gas and grain temperature,
and sticking factor.  Here, we express the grain formation rate per H atom
as a constant $R$ (in cm$^3$ s$^{-1}$), which is a free parameter in the
models; that is, we suppress explicit independent variation of all the
different quantities that affect the overall rate.  The volume rate of formation on grain surfaces is
then $n_{H}n_{HI} R$, where $R \approx 3 \times 10^{-17}$ cm$^3$ s$^{-1}$
for Galactic conditions (Hollenbach, Werner, \& Salpeter 1971; Jura 1974).

\section{Determining physical parameters} 
\subsection{Overview}
In the sections that follow, we apply our code to the
problem of determining how changes in physical conditions along a
line of sight affect observed column densities.  To do so, we have created
a large ($>$ 5000 element) grid of models of varying temperature, density, size,
UV radiation field, and grain formation rate of H$_2$.  Here, we examine how
changes in these major physical parameters are manifested in various
diagnostics, and we give scaling relations which approximate the
quantitative changes in molecular fraction incurred by altering the
radiation field or grain formation rate.  

\subsection{Diagnostics}
Many different quantities might in principle act as comparators between the
data and attempts to model them.  The measured column densities themselves
can of course be compared to calculated ones, but they do not by themselves
serve as reliable indicators of physical parameters like density,
temperature, and UV radiation field; direct comparison of column densities
is also unwieldy for large datasets.  In this section, we give a brief overview
of the additional diagnostics we have chosen, and the reasons why we have chosen them.

One such diagnostic is the fraction of hydrogen nuclei in molecular form,
\begin{equation}
f_{H2} = \frac{2 N(H_2)}{[N(HI) + 2N(H_2)]} .
\end{equation}
\noindent  In the standard simple model of an interstellar cloud (Jura 1975a,b), as
described above and implemented in our code, H$_2$
is assumed to be in equilibrium between formation on grains with rate
coefficient
$R$ (cm$^3$ s$^{-1}$) and photodestruction with rate $D$ (s$^{-1}$),
\begin{equation}
D n(H_{2}) = R n_{H} n_{HI} \approx 0.11 \sum_{J} \beta (J) n(H_{2},J) .
\end{equation}
Here, we have assumed that the fraction of absorptions which lead to
dissociation, $f_{diss}$, is 0.11.  In the code, we do not make this
assumption, but explicitly calculate $f_{diss}$ for each band. It varies
somewhat, depending on the radiation field and depth into the cloud, but
always lies in the range 0.10 -- 0.15.  

If one assumes that the cloud is homogeneous, with physical densities 
replaced by column densities, then one may relate $f_{H2}$, as defined above, to the
formation and photodestruction rates :
\begin{equation}
f_{H2} = \frac{2 R n_H}{D} .
\end{equation}
Thus, the molecular fraction can be suppressed either by a depressed
formation rate or an enhanced photodestruction rate.  Qualitatively, $f_{H2}$
acts as measure of the balance between these two processes.

To probe the temperature of the gas, one typically turns to the excitation
temperature describing the populations of $J=0$ and $J=1$, given
by
\begin{equation}
T_{01} = \frac{(\Delta E_{01}) / k}{\ln[(g_{1}/g_{0}) N(0)/N(1)]},
\end{equation}
where $g_{1}/g_{0} = 9$ is the ratio of statistical weights of the $J=1$
and $J=0$ levels, and $(\Delta E_{01})/k = 171$ K.  At densities high enough
for collisions with H$^{+}$ and H$_{3}^{+}$ to dominate the (0,1) level populations (such as in the
translucent clouds), $T_{01}$ must trace the kinetic temperature. At the
densities typical of diffuse clouds, the relation between these two
quantities is not precisely known (Tumlinson et al. 2002), but $T_{01}$
probably still traces $T_{kin}$ to some extent for $N(H_2) \geq 10^{16}$
cm$^{-2}$.  The kinetic temperature is a free parameter in our models.

One can construct similar excitation temperatures, or equivalent
quantities, for any two levels. However, these are decreasingly sensitive
to the kinetic temperature and increasingly sensitive to details of the
radiative cascade as one goes to increasing $J$.  In particular, we employ
the column density ratios N(3)/N(1), N(4)/N(2), and N(5)/N(3). The first
ratio is probably still sensitive to collisional processes in the gas (Jura
1975b; Tumlinson et al.  2002), but the latter two ratios likely probe the
radiation field rather than the kinetics of the gas.

One caveat to the use of column density ratios is that, while these ratios
can be a useful diagnostic of the radiation field (and potentially of the
thermal structure along a line of sight), they do represent a loss
of information relative to the individual column densities.   It is
possible for two sightlines to have very similar column density ratios
but very different actual column densities.

By examining all of these diagnostics together, we can assess
quantitatively the overall abundance of H$_2$ and its excitation, in both
models and data.  

\subsection{Thermal effects}
In Figure 3, we show two sets of models with a range of densities and
sizes (such that $N(H_2)$ $< 9.0 \times 10^{21}$ cm$^{-2}$), but with constant (and
high) UV radiation field, $I = 5 \times 10^{-7}$ photons cm$^{-2}$ s$^{-1}$
Hz$^{-1}$, about 50 times the Galactic mean. The models shown as triangles are at a
kinetic temperature $T=10$ K, while those shown as diamonds are at $T=90$ K.  Also shown
are ``trend lines'' connecting corresponding points in the two models;
these indicate how points in the excitation diagram move in response to
decreasing temperature.  In general, decreasing the kinetic temperature for
clouds with $N(H_2)$ $\gtrsim$ 10$^{20}$ cm$^{-2}$ moves points
towards higher excitation ratios N(4)/N(2) and N(5)/N(3).  This is largely a result of the fact that, at these
temperatures, $J=4$ and $J=5$ are not populated to a signficant extent
by collisional processes, but rather by radiative cascade.  Although
collisions indirectly modify the $J=4$ and $J=5$ populations, this effect
is small. The population of
$J=2$, however, is more directly affected by
collisional processes, albeit on a still-minor level. Thus, as the kinetic
temperature drops, a population source for $J=2$ disappears, while the
population sources of $J=4$ and higher do not.  This effect is only
noticeable when collisional effects are significant relative to radiative
ones for $J=2$, so at low column densities, where incident radiation is less
shielded, changing the temperature has little effect on the
excitation ratios.  At intermediate column densities, where collisional
population of $J=2$ is negligible but collisional population of $J=1$ is not,
changes in temperature can have the opposite effect on excitation.  As the
temperature rises in this regime, $J=1$ is populated at the expense of $J=0$,
thus reducing slightly the population rate via radiative cascade for both
$J=2$ and $J=4$. The small difference in physical density affects the column
density of $J=2$ more strongly than it does $J=4$, since the former level is
beginning to self-shield while the latter does not.  Thus, for column
densities $N(H_{2})$ between roughly 10$^{17}$ cm$^{-2}$ and 10$^{20}$
cm$^{-2}$, increases in kinetic temperature translate to increases in the
excitation ratio N(4)/N(2).

A minor complication in these analyses is that changing the temperature can
slightly affect the total $N(H_{2})$.  As the temperature drops, all
other things being equal, more material will reside in the lowest
rotational level, $J=0$.  This means that $J=0$ will become strongly
self-shielding earlier, depressing the photo-destruction rate further into
the cloud, and the total abundance of H$_2$ will rise slightly.  This effect is
only important in the critical regime where absorption out of $J=0$ is not
already highly damped; at higher column densities, a small increase in the
population of $J=0$ has no appreciable affect on the total photodestruction
rate and, hence, on the total molecular abundance.

\subsection{UV radiation field}

As the UV radiation field becomes stronger, the photo-destruction rate
clearly also rises; hence, a first-order effect of the enhanced radiation
field is to lower the molecular fraction.  This effect is displayed in the
top left panel of Figure 4, which shows molecular fractions for models at the
same density and temperature but different radiation fields.  As seen here,
an enhanced radiation field also has the effect of delaying the onset of
very high molecular fraction, since the higher rate of UV absorption means
that the major absorption lines out of $J=0$ and $J=1$ become optically
thick at larger $N_H$.  Thus, the critical regime where small changes in
column density correspond to large changes in $f_{H2}$ also occurs at
higher $N(H_2)$. A high UV radiation field can also change fairly
dramatically the ratios of the populations in each rotational-vibrational
level.  The bottom left panel of Figure 4 shows the excitation ratio N(4)/N(2) for
models at the same temperature ($T = 60$ K) and density ($n_H$ = 100
cm$^{-3}$), but with two different UV radiation fields.  Models in green
are exposed to a field roughly the Galactic mean, $I = 1 \times 10^{-8}$ photons
cm$^{-2}$ s$^{-1}$ Hz$^{-1}$; models in red are exposed to a field 50 times
as intense.  At lower radiation field intensities, a disparity quickly
arises between levels $J=0$ and $J=1$ on the one hand, and all higher
levels on the other. Because these two lowest levels have no means for
spontaneous de-excitation, absorption lines out of those levels become
damped. The rates out of these levels deep in the cloud become smaller by a factor $\sim
100$ compared to values at the edge of cloud.  Higher levels, on the
other hand, have spontaneous deexcitation probabilities that can quickly
become the dominant means for depopulating the level. The primary
population mechanism (radiative absorption and subsequent cascade into the
given level) is not strongly biased towards any one level or set of
levels. Thus, $J=0$ and
$J=1$ become a ``sink'' for H$_2$ relative to the other levels, with the
relative magnitudes of the level populations set largely by the ratio of
the dominant depopulation mechanisms (spontaneous radiative decay for $J
\geq$ 2, highly damped absorption for $J=0$ and $J=1$).

\subsection{Grain formation rate}

Changes in the grain formation rate also clearly affect the
molecular fraction. As less H$_2$ is formed on grains, all other things
being equal, the molecular fraction must drop.  In the top right panel of Figure 4, we show the
effects on molecular fraction of reducing the grain formation rate coefficient of H$_2$ by a
factor of ten; in this figure, all clouds are exposed to a Galactic mean
radiation field and have the same temperature ($T = 90$ K).

The effect on excitation ratios is, again, more subtle.  In the
bottom right panel of Figure 4, we show the excitation ratio N(4)/N(2) for models
exposed to the same Galactic mean radiation field, but with grain formation
rates which differ by a factor of ten.  Changes in
the grain formation rate can affect the ratios of column densities in each
ro-vibrational level by changing the total column density of H$_2$ for a
given physical density, radiation field, and temperature.  This change leads to more or less damping of absorption in the $J=0$ and $J=1$ Lyman and Werner
bands, and can therefore alter the level populations.  In general,
decreasing the grain formation rate enhances the excitation ratio
N(4)/N(2), but does not appreciably alter the ratio N(5)/N(3).  

As before, changes in both the total molecular abundance and in the
excitation ratios are most pronounced for clouds that lie on the cusp of
self-shielding (i.e., those with $N(H_{2})$ $\sim 10^{18}$ cm$^{-2}$).  

\subsection{Scaling relation}
\label{section:3.6}
Here, we give an approximate relation for the molecular fraction in a cloud
of H$_2$ as a function of the total hydrogen column density, $N_H$, the grain
formation rate of H$_2$, and the intensity of the UV radiation field.  This
relation is intended to be most accurate for conditions approaching nominal
Galactic ones; for environments with radiation fields or grain formation
rates far away from the Galactic ones, it is only a rough approximation.
In general, we consider a relationship of the type
\begin{equation}
\log(f_{H2}) = A_{0} q\left[1 - 0.98 \tanh\left(\frac{\log(N_{H}) -
A_{1} q}{A_{2} q}\right)\right] ,
\end{equation}
where 
\begin{equation}
q  =  \alpha(R)\beta(I) = \left(\frac{R}{R_{0}}\right)^{A_{3}}
\left(\frac{I}{I_{0}}\right)^{A_{4}} ,
\end{equation}
and we take $I_{0} = 10^{-8}$ photons cm$^{2}$ s$^{-1}$ Hz$^{-1}$ and
$R_{0} = 3 \times 10^{-17}$ cm$^{3}$ s$^{-1}$.
This functional form was chosen because it has the property of going from a
low value ($\sim 2 A_{0}$, where $A_{0}$ is a negative number) to a
high one ($\sim 0.98$, equivalent to the highest molecular fraction reached
by model clouds in this sample) at a characteristic column density set by the value
of $A_{1}$ and over a scale set by $A_{2}$.  The parameter $q$, a proxy for
the two independent parameters $A_3$ and $A_4$, expresses
variation in $R$ and $I$.  Including the variations with
respect to $R$ and $I$ makes this a four-parameter fit; allowing the
dependence on $R$ and $I$ to vary in each of the terms in this equation
does not improve the fit.

The best-fit
values were obtained by performing multi-dimensional parameter
minimizations against the results derived from the model cloud grid.  We
find $A_{0} = -2.4054$, $A_{1} = 20.178$, $A_{2} =
0.279$, $A_{3} = -0.0118$, and $A_{4} = 0.0124$.  In the leftmost panel of Figure 5, we show
model clouds with roughly Galactic radiation field ($I_{0}$) and grain
formation rate ($R_{0}$)
together with this scaling relation; the agreement here is good at all
points.  Some small vertical scatter in the molecular fractions of model
points is caused by the spread of temperatures in the models and
accompanying small changes in the molecular content of the gas (discussed
above). We have not included any variation with respect to temperature in
the scaling relation given here, because it
is generally much smaller than the effects induced by changing other
quantities.  In the middle panel of Figure 5, we show models exposed to a Galactic radiation
field but at a range of grain formation rates down to one-tenth the
Galactic mean.  The agreement with the scaling relation is generally
good, but begins to degrade at low molecular fractions and low $N_H$.  In the
rightmost panel of Figure
5, we include models exposed to a radiation field of up to $I = 1 \times
10^{-7}$ photons cm$^{-2}$ s$^{-1}$ Hz$^{-1}$.  Here, the agreement with the
data is fairly poor at very low molecular fractions ($f_{H2} \lesssim
10^{-5}$) but good at higher
ones; the scaling relation given here overestimates the molecular fraction
in highly optically thin clouds.  This problem persists at even higher
radiation fields.  Thus, this scaling relation cannot be used as a
substitute for detailed modeling in environments far removed from the
nominal Galactic one. It can, however, give an order-of-magnitude estimate
for the molecular fraction in such environments, and under Galactic
conditions it can be used in a more precise fashion.

We can find no such simple scaling relation for the excitation ratios
N(4)/N(2) and N(5)/N(3); too many effects combine to yield the observed
ratios for them to be easily described throughout the parameter space
considered here.  

\section{The translucent cloud paradigm}

In general, while the simple models we have presented here can duplicate
most diffuse cloud observations, they cannot match the observed populations
of very high-$N(H_{2})$ clouds ($N(H_{2})$ $\gtrsim 10^{20}$ cm$^{-2}$) without supposing
rather high incident UV radiation fields ($\gtrsim 20 \times $ Galactic mean).
This may be appropriate in some cases, but probably not all. In addition,
some of the observed high-column cloud populations cannot be explained by
the isothermal slab models at all.  In a single slab, it is difficult
to populate the rotational states $J > 4$ to the levels observed in these
targets without vastly overpopulating the lower-$J$ states.

We suggest that some high-$N(H_{2})$ clouds are combinations of
lower-$N(H_{2})$ diffuse clouds viewed along a common line of sight.  We do
not refer here to the idea of having a ``shell'' model for a cloud, wherein
a single line of sight consists of a monolithic object with one hot
exterior region and one cool interior.  Some lines of sight may indeed have
this sort of structure, but we also suggest the possibility of physically
distinct clouds, each exposed to an incident radiation field that is not
filtered through the other component clouds.  This interpretation is
appealing in part because of the results from single-component analyses, in
which high radiation fields are necessary to match any of the
high-$N(H_{2})$ line of sight observations.  If intense incident radiation
fields are not a possibility for a given line of sight, then there must be
additional pathways by which UV radiation is entering the system
(c.f. Rachford et al. 2001).  For
a given incident radiation field, one way of enhancing the effect of
that radiation field (and hence bringing the model clouds into agreement
with observations) is to allow it to enter the system at multiple points,
so it is not completely attenuated far into the cloud.

In Figure 6, we show H$_2$ observations along three selected high-$N(H_{2})$ lines of
sight : HD 108927 and HD 96675 (Gry et al. 2002), and HD 110432 (Rachford et
al. 2001).  For each line of sight, two model calculations are also shown
-- one that is the best single-cloud match to the data, and another, which
is a concatenation of two clouds.  Out of the many possible concatenations
that can match the data, we have chosen the best match for which both
component clouds are illuminated by a radiation field of no more than $4
\times 10^{-8}$ photons cm$^{-2}$ s$^{-1}$ Hz$^{-1}$, about four times the
Galactic mean.  In these two-cloud models, one cool component contributes
most of the total column density of H$_2$, while the other component,
smaller and hotter ($\sim 200$ K), helps populate the high-$J$ states.  The single-cloud
matches are all exposed to a high radiation field ($\gtrsim 2 \times
10^{-7}$ photons cm$^{-2}$ s$^{-1}$ Hz$^{-1}$), not because we have
excluded lower radiation field matches but because all individual models in our
database that can match the observed populations are exposed to such high
fields.

We have thus far been wary of describing these three sightlines as
``translucent clouds,'' because only two out of the three (HD 110432 and HD
96675) meet the defining requirement that the visual extinction $A_{V} > 1$.  (The
remaining target considered here, HD 108927, has $A_{V} \approx 0.68 \pm
0.1$.)  While the visual extinction may indeed be an important
discriminator of cloud properties (Rachford et al. 2001), there does not
seem to be a crucial difference between clouds with $A_{V} = 1-2$ and
those with slightly lower extinction.  All three clouds share the common
property that multiple radiation pathways (or a very high radiation field)
are required to explain their observed populations.  A more fundamental
distinction between ``diffuse'' and ``translucent'' clouds may be one of
molecular fraction: in models of all three sightlines, one component cloud
must have a fairly high molecular fraction ($f_{H2} \gtrsim 0.5$), in sharp
contrast to the very low molecular fractions ($f_{H2} \sim 10^{-5}$)
observed in many model clouds with $\log(N_{H}) \lesssim 21.25$ (for
Galactic grain formation rate and UV radiation field).  One difficulty with
such a definition, however, is that the molecular fraction is generally not
known along sightlines with high $N(H_{2})$; to estimate it, one uses an
assumed relationship between the observed color excess and the total
N$_{H}$. We also note that very high extinction clouds with A$_V$ $\gtrsim 5$
may be very different from the clouds considered here, but such
high-extinction clouds cannot be studied by FUSE because of the low FUV
fluxes of background stars.

Another high-$N(H_{2})$ sightline, towards HD 73882 (Snow et
al. 2000; Ferlet et al. 2000), is not well-matched by a two-cloud concatenation where neither
absorber is exposed to a high radiation field as described above.  It is
also not well-matched by any single cloud in our model database, regardless
of radiation field.  For these reasons, we have not included it in Figure
6.  It is possible that the absorbers towards HD 73882 (spectral type O8.5)
are exposed to a
high radiation field (Rachford et al. 2001), and the complicated component structure observed along the line
of sight makes it likely that the absorption is due to more than one
absorber.  Snow et al. (2000) note five distinct components along this
line of sight.  However, even allowing models with both high radiation
field and two components, we are unable to match the data.  This may be
because more than two components along the line of sight contribute heavily
to the H$_2$ absorption, a possibility not considered here because it is
computationally too intensive, or because some of the measured column densities
$N(J)$ may be incorrect.  If the population in $J=5$ is actually 10 times
lower than reported, and if the published population in $J=4$ is also higher
than the true one, the sightline can be matched by two-cloud models exposed
to a high radiation field; other errors in the measured $N(J)$ could also
lead to better agreement with the models.  Finally, we cannot rule out the
possibility that this sightline is fundamentally different from the others
considered here. It has a higher extinction, $A_{V} = 2.44$, than almost any
other line of sight observed in the UV (Rachford et al. 2002; Snow et al. 2000), and may probe a
different cloud regime than the other targets.  Once more targets at similar $A_{V}$ have been
observed, it may be possible to determine which of these different
scenarios is the case.  

There is circumstantial evidence that concatenations may be common outside
the Milky Way.  In a
recent survey of H$_2$ in the LMC and SMC, Tumlinson et al. (2002) found
that $92 \%$ of sightlines in the SMC (24 of 26) have $N(H_{2})$ $\gtrsim$
10$^{15}$ cm$^{-2}$. Approximately $23\%$ have $N(H_{2})$ in the range
$10^{14.5} - 10^{15.5}$ cm$^{-2}$, $23\%$ lie in the range $10^{15.5}
- 10^{16.5}$ cm$^{-2}$, $15\%$ are between $10^{16.5}$ and $10^{17.5}$ cm$^{-2}$,
$8\%$ are in the range $10^{17.5} - 10^{18.5}$ cm$^{-2}$, $19\%$ are
between $10^{18.5}$ and $10^{19.5}$ cm$^{-2}$, and $4\%$ have $N(H_{2})$ $>
10^{19.5}$ cm$^{-2}$.  Given that a low-$N(H_{2})$ component ($N(H_{2})$ $= 10^{14.5}
- 10^{16.5}$ cm$^{-2}$) is seen towards almost half of all the SMC targets,
we may conclude that such low-column clouds are common.
We then see no compelling reason to think that targets that show higher
$N(H_{2})$ along the line of sight do not also contain low
column-density absorbers.

\section{Model degeneracy}

 It is only by careful consideration of all the physical parameters that
affect the H$_2$ in a cloud (density, temperature, UV radiation field,
grain formation rate, etc.) that we can highlight the classes of models
that fit the observations.  The true nature of the lines of sight we study
is probably a blend of many of these effects. When one allows for the
possibility of multiple clouds contributing to the observed column
densities, the number of different scenarios that can yield a set of
observables becomes enormous.  In the absence of independent constraints on
the component structure or the physical properties of the absorbing medium,
it is not possible uniquely to identify which of these possible scenarios
is the true one.  In Figure 7, we illustrate a set of models exposed to a
high radiation field (10$^{-7}$ phot cm$^{-2}$ s$^{-1}$ Hz$^{-1}$) but of
varying density and size, together with a region commensurate with the
error bars on a measurement of H$_2$ towards AV 47 in the SMC (Tumlinson et
al. 2002).  Models shown as diamonds are at 120 K; models shown as circles
are at 10 K.  Multiple models from these disparate temperature regimes fall
within the region permitted by the error bars.

One criticism that might be leveled at our claim of non-uniqueness is that
the model degeneracy could disappear if we had included all other
potentially relevant molecular and atomic species in our calculation, and
had observations of those species.  While including sophisticated chemical
reaction networks in our models would in principle yield more observable
quantities (the column densities of each molecular species), it would also
add new parameters (the abundances of each species).  We therefore think it
unlikely that including a large number of chemical species would completely
remove any modeling degeneracy.  It is possible, however, that careful
modeling of a few additional species sensitive to the same UV photons as
H$_{2}$, coupled with observations of same, might provide additional
constraints; the degeneracies noted here might therefore be considered as
a worst-case scenario. We also note that for low-$N(H_{2})$
diffuse clouds, complicated chemistry networks probably play a minimal role
in determining cloud structure; the abundances of species other than
hydrogen are simply too low.

\section{FUSE survey of H$_2$ in the LMC and SMC}
While it may not be possible to uniquely constrain the physical conditions
along a single line of sight, our models can still be applied to
an ensemble of sightlines to yield useful information.  If all models that can
match a given dataset have some traits in common, those traits might
reasonably be assumed to be present in the real clouds as well. Likewise,
if all of the models with a given set of properties fail to describe
the data, these properties are probably absent from the real clouds.  It is
with these considerations in mind that we apply the models to a FUSE survey
of H$_2$ in the LMC and SMC.

The survey, described fully in Tumlinson et al. (2002), consists of 70
sightlines containing H$_2$ in the LMC and SMC. The resulting data were
compared with a grid of 3780 models.  This grid consisted of
isothermal slabs (as described above) of varying temperature ($T$ =
10, 30, 60, 90, 120, 150 K), density ($n_{H}$ = 5, 25, 50, 100, 200, 400,
800 cm$^{-3}$), size ($d$ = 2, 4, 6, 8, 10 pc),
mean UV radiation field ($I$ = 1.0, 4.0, 10, 20, 50, 100 $\times 10^{-8}$
photons cm$^{-2}$ s$^{-1}$ Hz$^{-1}$), and
grain formation rate coefficient ($R$ = 0.3, 1.0, 3.0 $\times 10^{-17}$  cm$^{3}$
s$^{-1}$).  The grid of models is labeled in Table 4.

In Figure 8, we compare the observed molecular fractions from the LMC, SMC,
and Milky Way samples, with our model clouds.  Model grid A, representing
typical Galactic conditions, and model grid D, which has both an enhanced
radiation field ($I = 10^{-7} - 10^{-6}$ photons cm$^{-2}$
s$^{-1}$ Hz$^{-1}$) and reduced grain formation rate coefficient ($R = 3
\times 10^{-18}$ cm$^{3}$ s$^{-1}$), are overlaid with the data.  Also
shown are lines connecting points in grids A and D with the same density,
size, and temperature, but different radiation fields and grain formation
rates. These lines illustrate the general trend, noted above, that
enhancing the radiation field and/or lowering the grain formation rate
tends to lower the molecular fraction as well as shift the characteristic
breakpoint in column density between gas that is strongly self-shielding
and gas that is not.  

While the models in grid A match the Galactic points (in blue) fairly well,
they do not overlap with the LMC or SMC points at all.  Model grids which
have either the enhanced radiation field or reduced grain formation rate of
grid D, but not both, are also generally overabundant in H$_2$ relative to
the data; these intermediate grids are not shown here.  Only the extreme case of
model D is able to match the abundance pattern of most of the LMC and
SMC points.  

In Figure 9, we examine the column density ratios, N(4)/N(2) and N(5)/N(3),
of the three samples, together with shaded regions indicating model points
and two-cloud concatenations.   The difference
between the excitation patterns observed in the LMC or SMC and that in the
Milky Way is readily apparent.  To try to explain it, we turn to the same
grids of models used in analyzing the molecular fraction data.  Again,
grids representing typical Galactic conditions cannot duplicate the
observed excitation pattern -- see Tumlinson et al. (2002) for a comparison
of individual model grids with the data.  The best match to the data is
achieved by models with both low $R$ and high $I$. 

There are still some LMC and SMC points whose excitation ratios cannot be
explained, even by models with enhanced $I$ and reduced $R$.  These points
may be concatenations of multiple clouds, as described above, although to
explain some of the N(4)/N(2) observations, one of the component clouds
must be quite hot ($T >$ 400 K).  We note that such hot clouds of H$_2$ may
not be possible, since H$_2$ formation may be strongly suppressed at
temperatures $T \gtrsim 200$ K (Shull \& Beckwith 1982).  In Figure 9, the
region shaded red and black indicates the area of the excitation plots that
can be reached by single-model clouds in grid D, discussed above; the
region covered only by red lines indicates points in the excitation plot
that can only be reached by a concatenation of two model clouds.  Much of
the data can be matched by such concatenations, but some observed N(4)/N(2)
ratios are still unexplained. The highest N(5)/N(3) ratios are achieved by
combinations in which one cloud is exposed to a relatively low radiation
field (roughly Galactic mean intensity), and another is exposed to a very
high radiation field (10 to 100 times Galactic mean). This is unsurprising,
since the highest N(5)/N(3) ratios are achieved by combinations of clouds
with the lowest population in $J=3$ (low radiation field) and clouds with
the highest population in $J=5$ (high radiation field).  Because a
concatenation of multiple clouds can duplicate an overall increase in
radiation field intensity, if all the lines of sight studied here are
actually composed of multiple components, the very high radiation field
suggested here would be unnecessary. However, it seems unlikely that such
concatenations would occur much more often in the Clouds than in the Milky
Way and thus explain all the observed populations.

A number of potential alternative processes do not suffice to explain the observed LMC/SMC
properties:
\begin{itemize}
\item The disparity between the LMC/SMC points and the Galactic ones
cannot be due solely to differing kinetic temperatures in the two
environments.  In the model grid with Galactic conditions, the
coldest clouds (with $T = 10$ K) show the highest N(4)/N(2) and
N(5)/N(3) ratios. Decreasing the kinetic temperature below this point is
physically unreasonable and has only a marginal effect; increasing
it only moves the model points farther away from the LMC/SMC data.

\item ``Formation pumping,'' the idea that H$_2$ formation may preferentially
populate high-J levels, does not suffice to explain the observed
excitation.  Because absorption in the Lyman and Werner bands is followed
by photodissociation only $\sim$ 11\% of the time, and because
formation must balance photodissociation rather than total absorption, the
formation distribution of molecules has a small effect on the final
column densities.  We have run sample calculations in which all
formation was into the $J=4$ and $J=5$ levels, yet even in this extreme
case the resulting level populations are not appreciably different from
those given here.  

\end{itemize}

We conclude, on the basis of the abundance and excitation data presented
above, that H$_2$ in the SMC and LMC is formed on grains at a rate
approximately 10-40 \% that in the Galaxy, and that some of the H$_2$ is exposed
to a radiation field 10-100 times more intense than the Galactic one.
Ensembles of models under those conditions, though not unique, can
reproduce the observed abundance and excitation patterns.  There is some
evidence that some of the LMC/SMC gas is exposed to a weaker (roughly
Galactic) radiation field; not all of the observed sightlines
display high H$_2$ rotational excitation, and as noted above, the best matches to 
the highest observed ratios come from combinations of clouds where one
component is irradiated by a very strong field and the other is exposed to
a weaker field.  There is probably a blend of environments for H$_2$
in the Clouds, whereby some of the gas is irradiated at levels not much
more intense than in the Milky Way.  However, a portion of the
H$_2$ must be exposed to the very high radiation fields noted here in order
to explain the observed excitations and molecular fractions.

The above ideas are qualitatively in keeping with the low dust-to-gas ratios previously deduced
for the Magellanic Clouds (Koornneef 1982; Fitzpatrick 1985).  Those ratios
imply a smaller grain surface area per H atom (hence lower $R$) and less
attenuation of incident radiation. 

\section{Conclusions}

We have presented computational models of clouds of H$_2$ and selected
applications of those models to FUSE observations.  Relatively simple
models can duplicate the observed properties of diffuse and translucent
clouds fairly well, but there are complications.

Changes in the major physical parameters of such clouds can have many competing effects on
observables, and we have demonstrated the general trends obeyed by changes
in each parameter.  The effects of changes in one parameter can be mirrored
by changes in others, so uniquely identifying a line of sight with one set
of physical properties is in some cases not feasible.  Rather,
consideration should be given to the classes of models which can fit
a given ensemble of observations.  

We suggest that some high-$N(H_{2})$ ``clouds'' may be separate and
physically distinct absorbers viewed along a common line of sight.  This
interpretation is supported by the fact that the modeled rotational excitation matches the observations,
but also by what is needed if one does not allow such combinations: the
high-$N(H_2)$ cloud targets can generally be matched by a single cloud only
if that cloud is exposed to a high ($\gtrsim$ 20 times Galactic mean)
radiation field.  For some Galactic targets, such a radiation field is
probably not likely; to explain these observations without recourse to high
radiation field requires some additional pathway by which radiation may
enter the system.  Multiple-component clouds represent such a pathway, if
each cloud is physically distinct from the others and therefore does not
filter out the incident radiation.

Finally, we have illustrated the application of the code to a large FUSE
survey of H$_2$ in the LMC and SMC.  Ensembles of models, though not
unique, can match the observed patterns of molecular abundance and
rotational excitation.  We find evidence for an enhanced UV radiation field
(10 to 100 times the Galactic mean), and a reduced grain formation rate of
H$_2$ ($R \approx 3 \times 10^{-18}$ cm$^{3}$ s$^{-1}$, one-tenth the
nominal Galactic rate).
\acknowledgements
We thank D.R. Flower, E. van Dishoeck, H. Abgrall, E. Roueff, and B. Draine
for providing electronic compilations of the relevant radiative and
collisional rates.  We also thank C. Gry and co-authors for allowing us to
use their results prior to publication, and B. Rachford for many helpful
discussions. This work was supported in part by the Colorado
astrophysical theory program through NASA grant NAG5-7262.  This work is
based on data obtained for the Guaranteed Time Team by the NASA-CNES-CSA
\emph{FUSE} mission operated by Johns Hopkins University.  Financial
support to US participants has been provided by NASA contract NAS5-32985.

\vspace*{2.5in}
\renewcommand{\arraystretch}{1.2}
\begin{deluxetable}{ccccccccc}
\tablecolumns{9}
\tablenum{1}
\tablecaption{Reference cloud\tablenotemark{a} \ model column densities (cm$^{-2}$)}
\tablehead{
\colhead{Level} & \colhead{C1} &  \colhead{vDB C1}
& \colhead{C4}  &  \colhead{vDB C4}
      & \colhead{C7}  & \colhead{vDB C7}  & \colhead{C10}
& \colhead{vDB C10}
}
\startdata
$J=0$ & 3.9$\times 10^{20}$ & 4.1$\times 10^{20}$ & 3.5$\times 10^{20}$
      & 3.7$\times 10^{20}$ & 2.5$\times 10^{20}$
      & 2.7$\times 10^{20}$ & 1.5$\times 10^{20}$
      & 1.6$\times 10^{20}$ \\
$J=1$ & 1.3$\times 10^{18}$ & 5.5$\times 10^{18}$ 
      & 4.5$\times 10^{19}$ & 5.0$\times 10^{19}$ 
      & 1.3$\times 10^{20}$ & 1.5$\times 10^{20}$ 
      & 2.4$\times 10^{20}$ & 2.6$\times 10^{20}$ \\
$J=2$ & 2.0$\times 10^{17}$ & 1.9$\times 10^{17}$ 
      & 2.1$\times 10^{17}$ & 1.8$\times 10^{17}$ 
      & 1.5$\times 10^{17}$ & 3.1$\times 10^{17}$ 
      & 6.7$\times 10^{17}$ & 2.8$\times 10^{17}$ \\
$J=3$ & 3.5$\times 10^{15}$ & 4.3$\times 10^{15}$ 
      & 5.8$\times 10^{15}$ & 3.9$\times 10^{15}$ 
      & 8.4$\times 10^{15}$ & 1.4$\times 10^{16}$
      & 1.2$\times 10^{16}$ & 3.5$\times 10^{16}$ \\
$J=4$ & 1.5$\times 10^{15}$ & 1.4$\times 10^{15}$
      & 1.5$\times 10^{15}$ & 1.4$\times 10^{15}$
      & 1.1$\times 10^{15}$ & 1.3$\times 10^{15}$
      & 8.8$\times 10^{14}$ & 1.2$\times 10^{15}$ \\
$J=5$ & 8.3$\times 10^{13}$ & 7.5$\times 10^{13}$
      & 1.2$\times 10^{14}$ & 1.4$\times 10^{14}$
      & 1.5$\times 10^{14}$ & 2.1$\times 10^{14}$
      & 1.9$\times 10^{14}$ & 2.8$\times 10^{14}$ \\
$J=6$ & 3.8$\times 10^{13}$ & 2.9$\times 10^{13}$
      & 3.5$\times 10^{13}$ & 2.9$\times 10^{13}$
      & 3.0$\times 10^{13}$ & 2.6$\times 10^{13}$
      & 2.7$\times 10^{13}$ & 2.5$\times 10^{13}$ \\
\enddata
\tablenotetext{a}{$n_{H} =$ 250 cm$^{-3}$, $I =$ 2 $\times
10^{-8}$ ph cm$^{-2}$ s$^{-1}$ Hz$^{-1}$, $R = 3 \times 10^{-17}$ cm$^{3}$
s$^{-1}$, $T =$ 20, 40, 60, 100 K.}
\tablecomments{Models with prefix ``vDB'' are from van Dishoeck \& Black
  (1987), and are labeled as in that text.  Models without a prefix are
  comparable to the corresponding model from vDB.}
\end{deluxetable}
\renewcommand{\arraystretch}{1.0}

\renewcommand{\arraystretch}{1.05} 
\begin{deluxetable}{cccccccc}
\tablecolumns{8}
\tablenum{2}
\tablecaption{One and two-sided slab models\tablenotemark{a}}
\tablehead{
\colhead{Label} & \colhead{N(0)} &   \colhead{N(1)}
& \colhead{N(2)}  &  \colhead{N(3)}
      &  \colhead{N(4)} & \colhead{N(5)}  & \colhead{N(6)} \\
& \colhead{(cm$^{-2}$)} & \colhead{(cm$^{-2}$)} & \colhead{(cm$^{-2}$)} &
\colhead{(cm$^{-2}$)} & \colhead{(cm$^{-2}$)} & \colhead{(cm$^{-2}$)} &
\colhead{(cm$^{-2}$)} }

\startdata
One-sided    & 2.8$\times 10^{20}$   & 8.1$\times 10^{18}$ & 5.2$\times 10^{18}$ & 7.9$\times 10^{16}$ & 1.3$\times 10^{15}$ &
3.4$\times 10^{14}$ & 4.3$\times 10^{13}$    \\ 
Two-sided    & 2.7$\times 10^{20}$  & 8.1$\times 10^{20}$ & 5.9$\times 10^{18}$ & 8.8$\times 10^{16}$ & 1.4$\times 10^{15}$ &
3.7$\times 10^{14}$ & 4.6$\times 10^{13}$  
\enddata
\tablenotetext{a}{$T =$ 150 K, $n_{H} =$ 400 cm$^{-3}$, $I =$ 4 $\times
10^{-8}$ ph cm$^{-2}$ s$^{-1}$ Hz$^{-1}$, $R = 3 \times 10^{-17}$ cm$^{3}$
s$^{-1}$}
\end{deluxetable}
\renewcommand{\arraystretch}{1.0}

\renewcommand{\arraystretch}{1.2} 
\begin{deluxetable}{cccccccc}
\tablecolumns{8}
\tablenum{3}
\tablecaption{Column densities for models\tablenotemark{a} \ exposed to flat and variable
radiation fields} 
\tablehead{
\colhead{Spectrum} & \colhead{N(0)} &  \colhead{ N(1)}
& \colhead{N(2)}  &  \colhead{N(3)}
      &  \colhead{N(4)} & \colhead{N(5)}  & \colhead{N(6)} \\
      & \colhead{(cm$^{-2}$)} & \colhead{(cm$^{-2}$)}
& \colhead{(cm$^{-2}$)} & \colhead{(cm$^{-2}$)} & \colhead{(cm$^{-2}$)} &
\colhead{(cm$^{-2}$)} & \colhead{(cm$^{-2}$)} }
\startdata
Flat    & 3.9$\times 10^{20}$   & 1.3$\times 10^{18}$ & 2.0$\times 10^{17}$ & 3.5$\times 10^{15}$ & 1.5$\times 10^{15}$ &
8.3$\times 10^{13}$ & 3.8$\times 10^{13}$    \\ 
Draine78   & 3.4$\times 10^{20}$  & 1.3$\times 10^{17}$ & 2.5$\times 10^{17}$ & 5.0$\times 10^{15}$ & 2.0$\times 10^{15}$ &
1.2$\times 10^{14}$ & 5.4$\times 10^{13}$   \\ 
\enddata
\tablenotetext{a}{$T =$ 20 K, $n_{H} =$ 200 cm$^{-3}$, flat $I =$ 2 $\times
10^{-8}$ ph cm$^{-2}$ s$^{-1}$ Hz$^{-1}$, $R = 3 \times 10^{-17}$ cm$^{3}$
s$^{-1}$}
\end{deluxetable}
\renewcommand{\arraystretch}{1.0} 

\vspace{1in}
\renewcommand{\arraystretch}{1.2}
\begin{deluxetable}{cccc}
\tablecolumns{4}
\tablenum{4}
\tablecaption{H$_2$ Cloud Model Grid}
\tablehead{
\colhead{Label} &  \colhead{R $(10^{-17})$}     &
\colhead{$I (10^{-8})$}  &  \colhead{Description}\\ 
      &  \colhead{(cm$^{3}$ s$^{-1}$)}
& \colhead{(ph cm$^{-2}$ s$^{-1}$ Hz$^{-1}$)} &
}
\startdata
A    & 1 - 3         & 1 - 4     &  Galactic conditions    \\ 
B    & 0.3           & 1 - 4     &  \\ 
C    & 1 - 3         & 10 - 100  &  \\ 
D    & 0.3           & 10 - 100  &  LMC, SMC \\ 
\enddata
\end{deluxetable}
\renewcommand{\arraystretch}{1.0}

\clearpage
\begin{figure}[hpt]
\center
\plotone{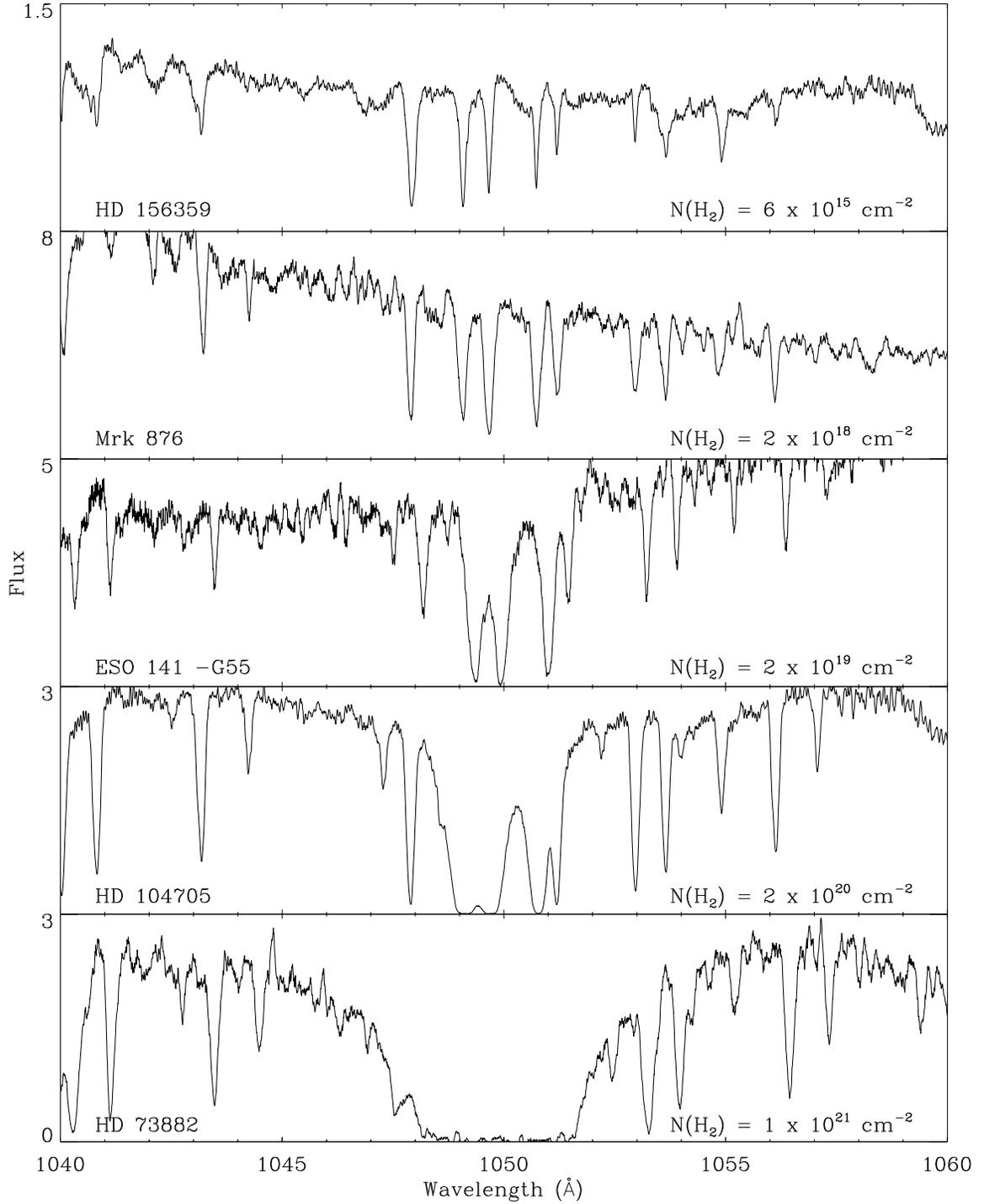}
\caption{FUSE observations of the (4-0) Lyman band for five sightlines in
  different $N(H_{2})$ regimes.  Note the effects of line overlap of the
  R(0), R(1), and P(1) lines (1049--1051 \AA) as $N(H_{2})$ approaches
  10$^{21}$ cm$^{-2}$.}
\end{figure}

\clearpage
\begin{figure}[hpt]
\center
\plotone{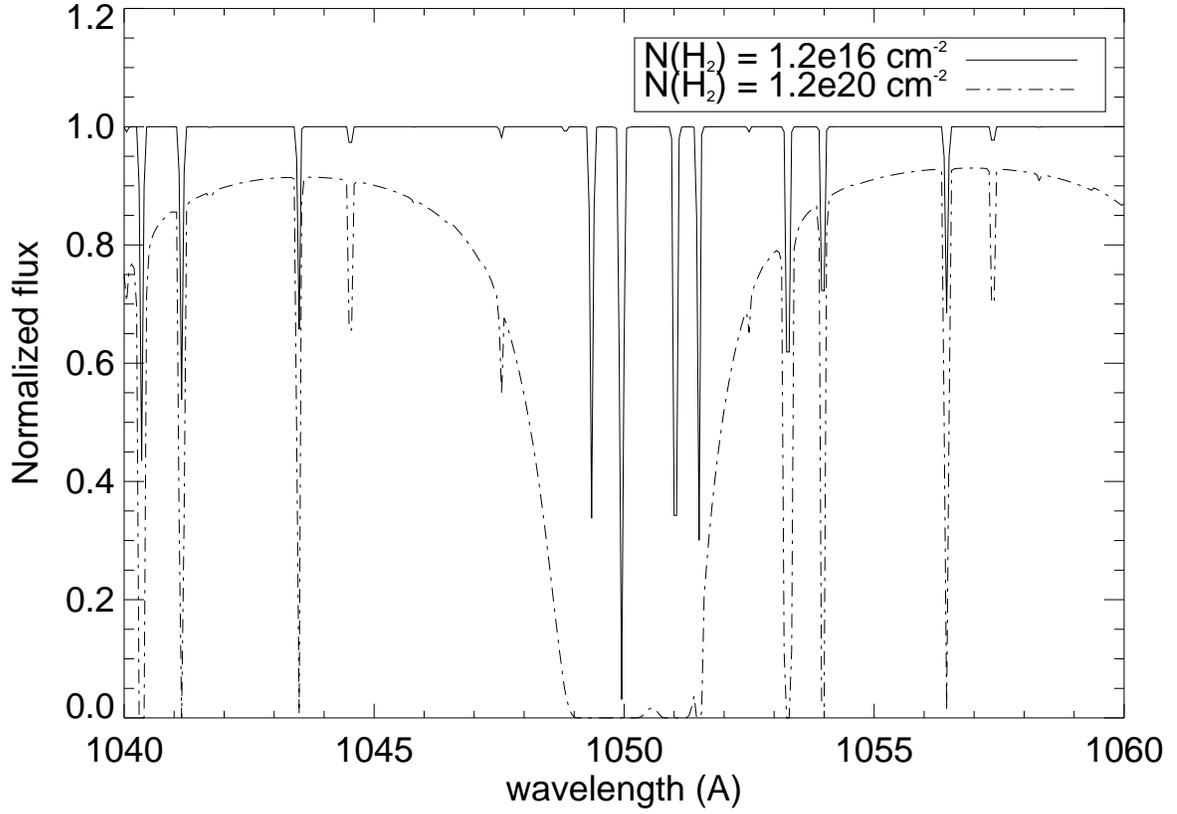}
\caption{Simulated spectra near the (4-0) Lyman band for a cloud of
physical density $n_{H}=70$ cm$^{-3}$ and temperature T = 98 K. Dashed-dot line
corresponds to $N(H_{2})$ = $1.2 \times 10^{20}$ cm$^{-2}$;
solid line corresponds to $1.2 \times 10^{16}$ cm$^{-2}$.}
\end{figure}

\begin{figure}[hpt]
\center
\plotone{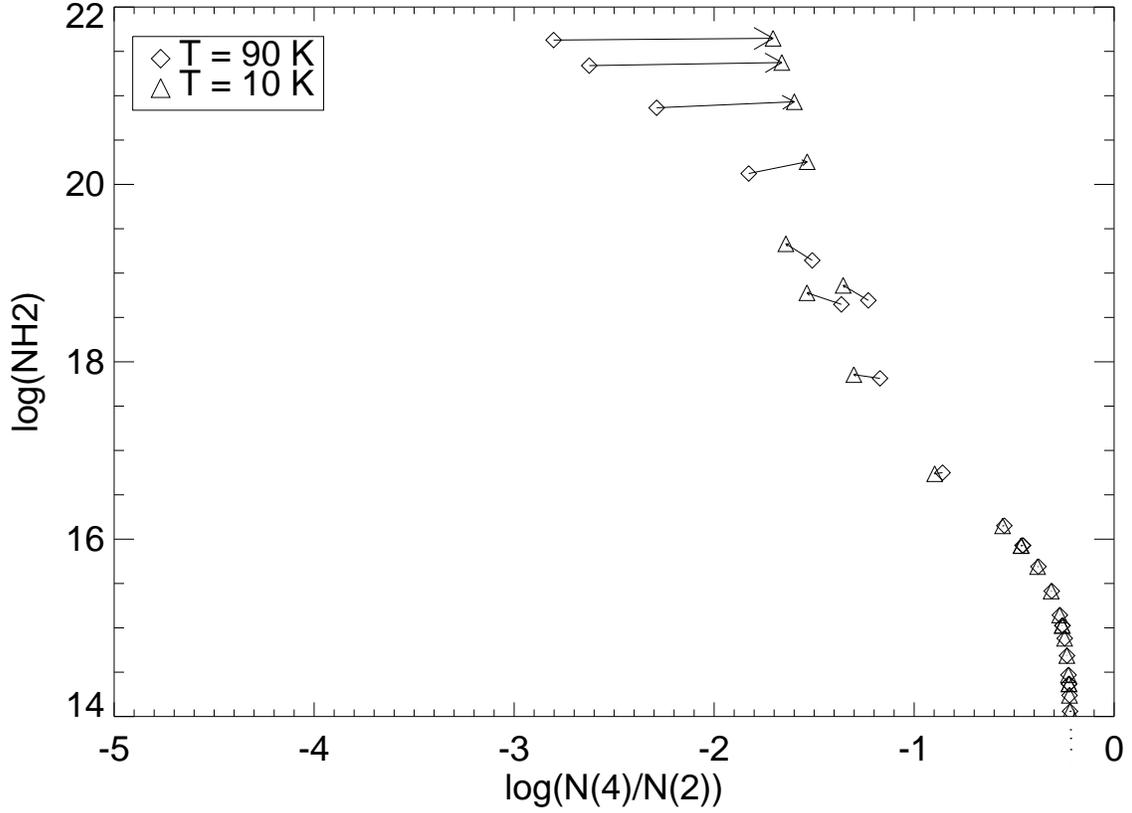}
\caption{Model clouds at $I=5 \times 10^{-7}$ photons cm$^{-2}$ s$^{-1}$
Hz$^{-1}$, but two different temperatures.  Lines are drawn between points
for which all properties except temperature are identical.  Points shown as
diamonds have $T=90$ K, while those shown as triangles are at $T=10$ K.  At
low column densities, changing the temperature has little effect because
collisional effects are unimportant relative to radiative decay.  The
behavior at higher $N(H_{2})$ is set by whether collisions are important in
populating $J=1$ and $J=2$.  See text for discussion.  }
\end{figure}

\begin{figure}[hpt]
\center
\plotone{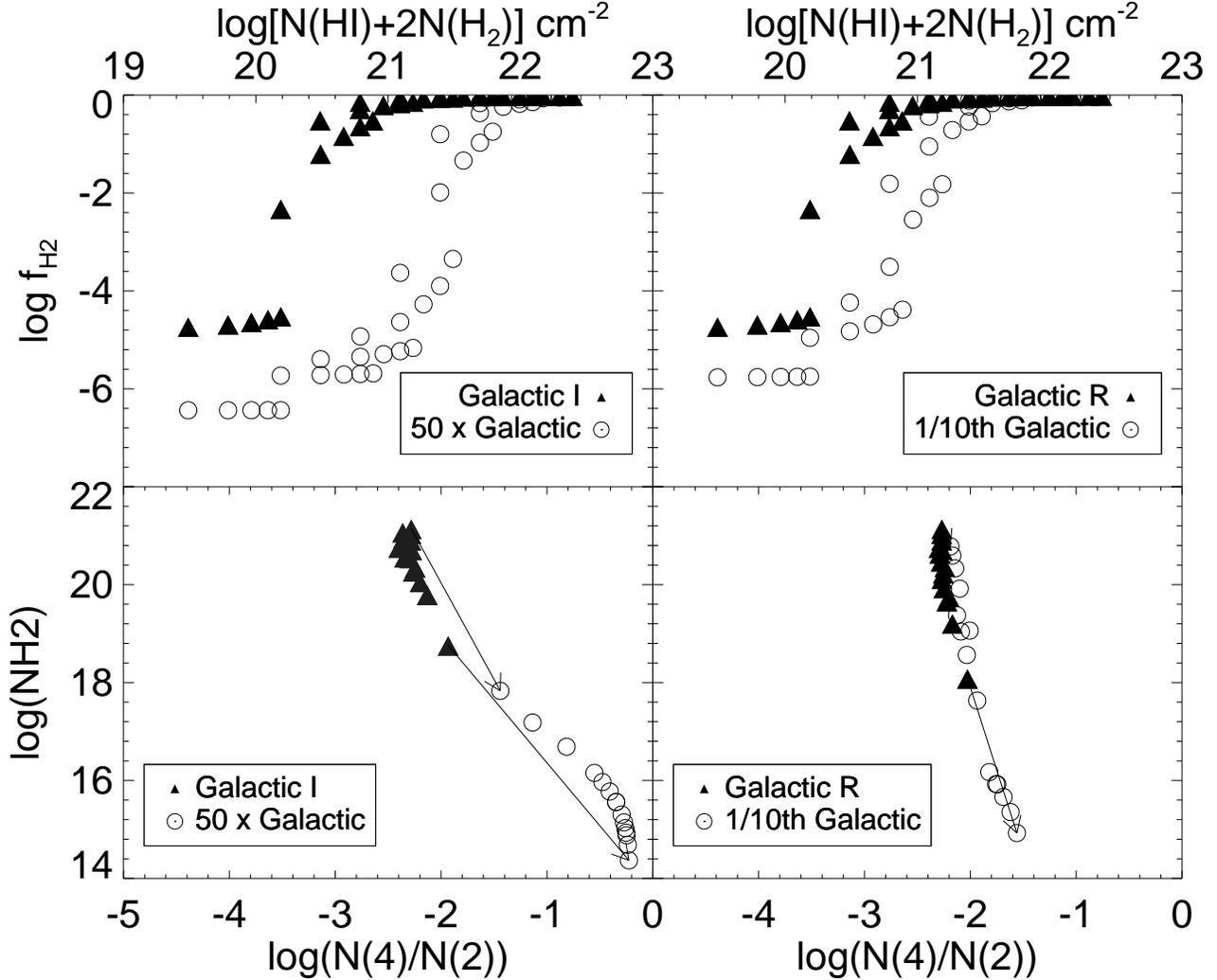}
\caption{Molecular fraction versus total hydrogen column density, and column
density of H$_2$ versus rotational excitation ratio in $J = 4$ and $J =
2$.  The left panels show model clouds exposed to Galactic mean UV
radiation field, together with models exposed to a field 50 times more
intense. The right panels show models at Galactic radiation field and grain
formation rate coefficient $R$, together with models with a grain formation
rate one-tenth the Galactic value.  Enhancing the radiation field or
decreasing the grain formation rate lowers the total abundance of H$_2$ and
pushes the transition to high molecular fraction towards higher column
densities. It also raises the excitation ratio N(4)/N(2).  The change in
excitation is more pronounced for changes in the radiation field than
for changes to the grain formation rate.}
\end{figure}


\begin{figure}[hpt]
\center
\plotone{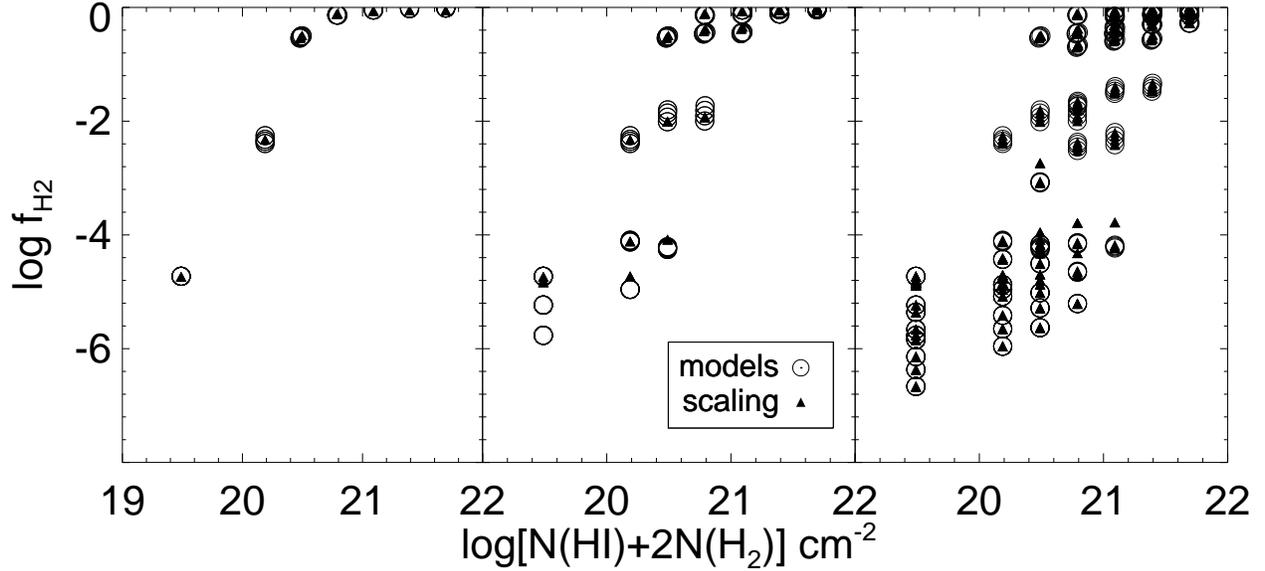}
\caption{Comparison of scaling relation for molecular fraction (see \S 3.6)
with models.  Models shown in far left panel have a range of densities, but
with grain formation rate and UV radiation field equal to the nominal
Galactic values.  Agreement between the models and the scaling relation is
quite good under these conditions.  Models in the middle panel have a range
of densities, with UV radiation field equal to the nominal Galactic value
and grain formation rate $R$ varying from $3 \times 10^{-18}$ to $3 \times
10^{-17}$ cm$^{3}$ s$^{-1}$.  Agreement between the models and the scaling
relation is fairly good, particularly at higher molecular fractions. Models
in the rightmost panel have a UV radiation field in the range $I =
(1-10) \times 10^{-8}$ photons cm$^{-2}$ s$^{-1}$ Hz$^{-1}$ and
grain formation rate $R$ varying from $3 \times 10^{-18}$ to $3 \times
10^{-17}$ cm$^{3}$ s$^{-1}$. Agreement between the models and the scaling
relation is fairly good at high molecular fractions but poor at the lowest
ones, corresponding to $I$ in excess of about $8 \times 10^{-8}$ photons
cm$^{-2}$ s$^{-1}$ Hz$^{-1}$.}
\end{figure}



\begin{figure}[hpt]
\center
\plotone{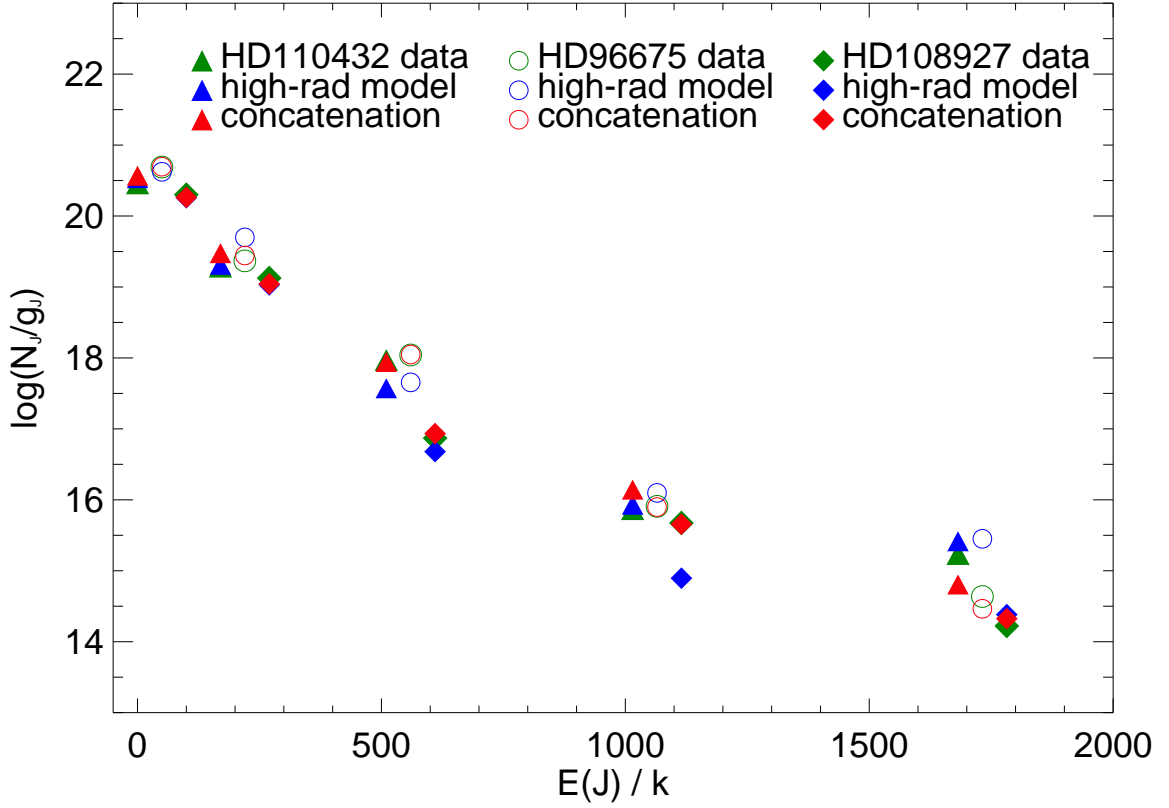}
\caption{Populations $N(J)$ vs. rotational energy E(J), for three high-$N(H_{2})$ sightlines, together
with both single-cloud and two-component models.  Each sightline is
represented by a different symbol; triangles correspond to HD 110432,
circles to HD 96675, and diamonds to HD 108927.  Green, blue, and red points
correspond, for each target, to data, high radiation field model, and
concatenation model respectively.  The sightlines have been
displaced slightly along the $x$ axis for clarity.  See text for discussion.
}
\end{figure}

\clearpage
\begin{figure}[hpt]
\center
\plotone{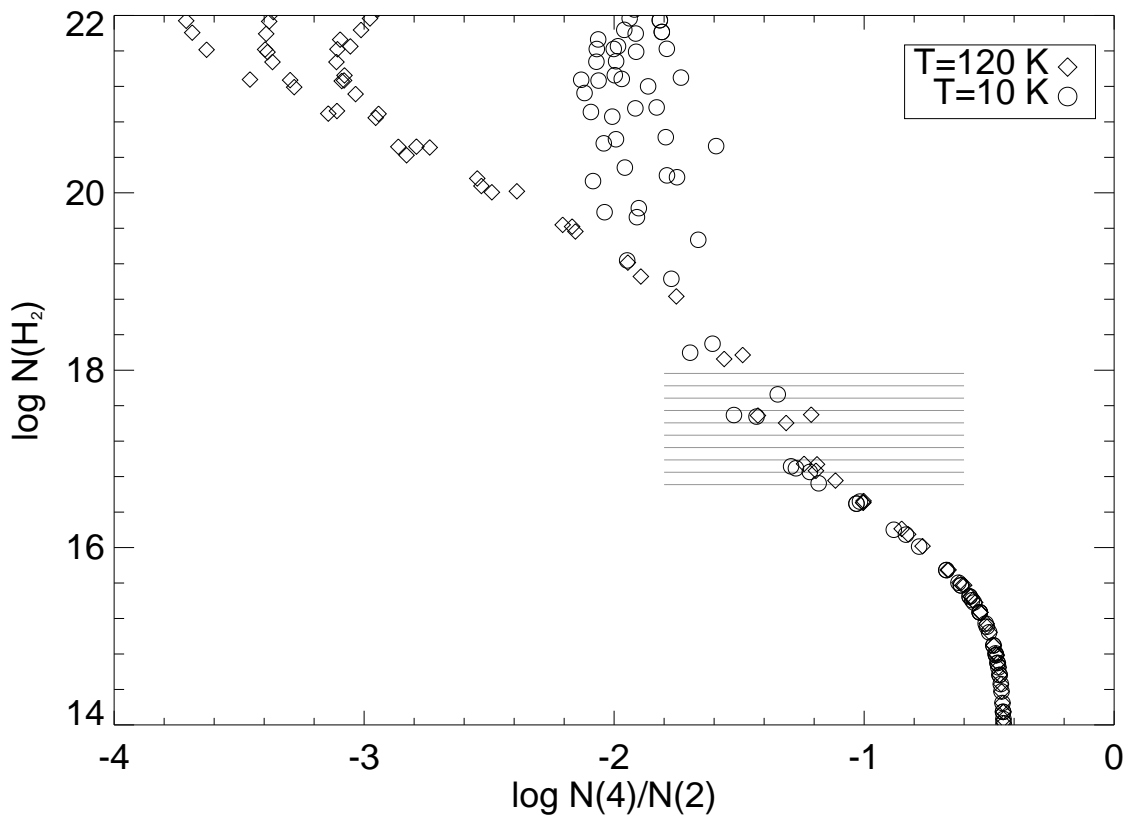}
\caption{ Column density $N(H_{2})$ in cm$^{-2}$ vs. ratio, N(4)/N(2), in
rotational states $J=4$ and $J=2$.  Models exposed to a high radiation
field (10$^{-7}$ phot cm$^{-2}$ s$^{-1}$ Hz$^{-1}$, $\sim$ 10 times
Galactic) but of varying density and size, together with a region
corresponding to the measurement and error bars associated with measurement
of H$_2$ towards AV 47 (designated by the bars). Models shown as diamonds are
at 120 K, models shown as circles are at 10 K.  Multiple models from these two
disparate temperature regimes fall within the region permitted by the error
bars.  }
\end{figure}

\begin{figure}[hpt]
\center
\plotone{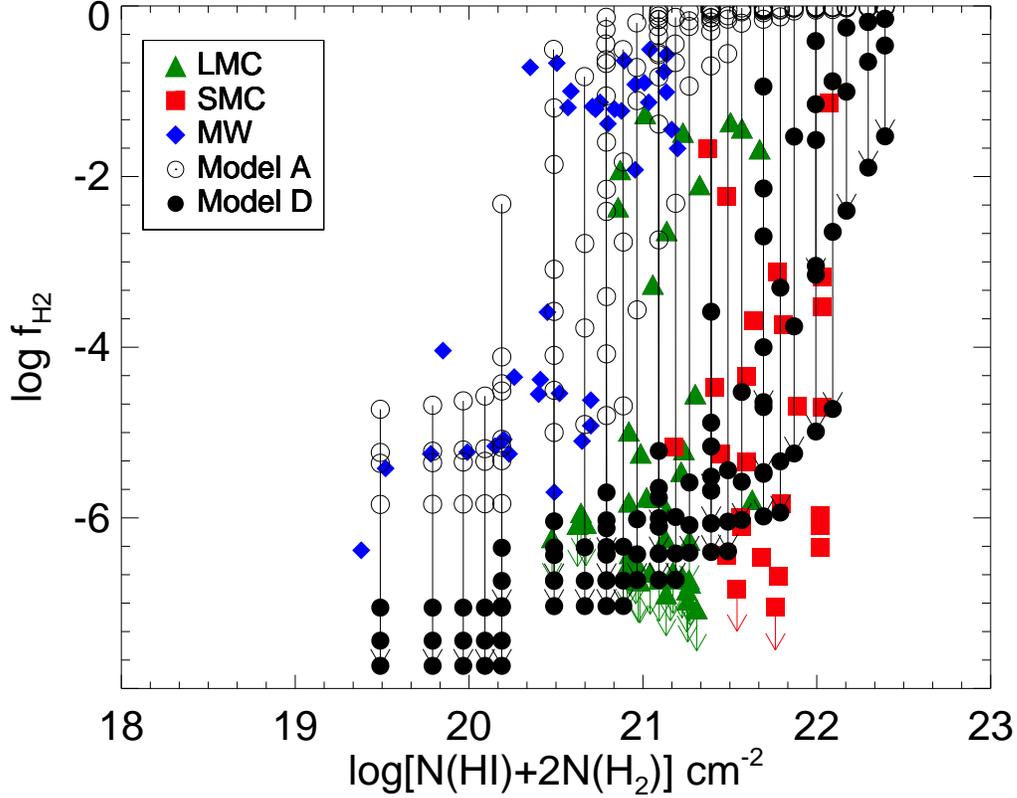}
\caption{Molecular fractions of SMC, LMC, and Milky Way samples plotted
against total H content and compared to model grids.  Models in grid A,
designated by open circles, have Galactic values of $R$, $I$, and the full
range of kinetic temperature, gas density, and cloud size.  Models in grid
D, designated by filled circles, have the same range of temperature, density, and size, but have $R = 3
\times 10^{-18}$ cm$^{3}$ s$^{-1}$, $\sim$ 0.1 times the Galactic rate, and $I =
10-100$ times the $\sim$ Galactic value.  Also shown are lines connecting points
in grid A to the correspoding points in grid D; connected points have all
properties except radiation field and grain formation rate in common.  The
agreement with the LMC and SMC data indicates enhanced radiation and reduced
grain formation rate of H$_2$ in the Clouds.  See text for more
discussion.}
\end{figure}

\begin{figure}[hpt]
\center
\plotone{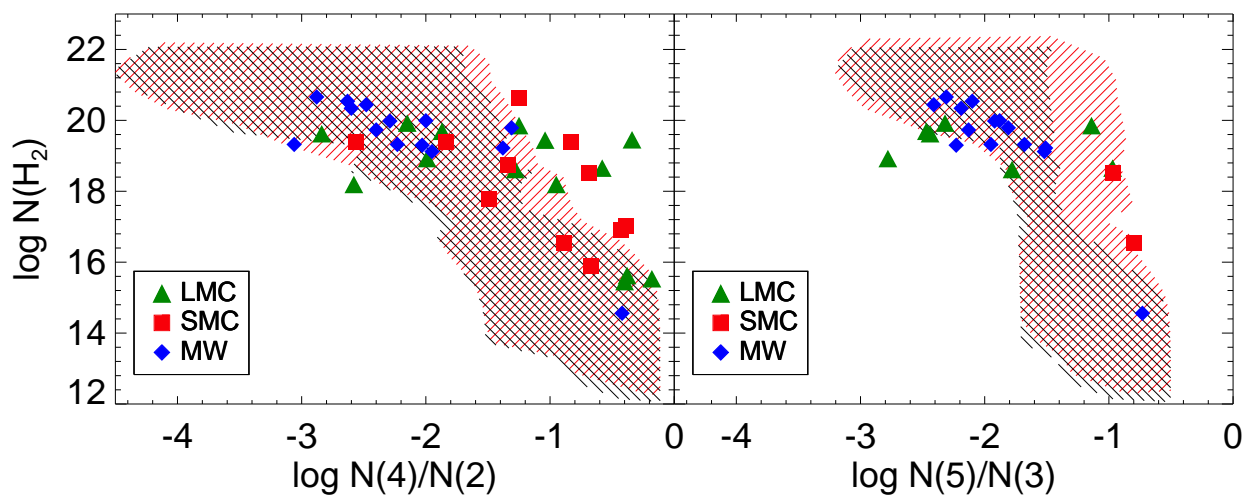}
\caption{Total column density of H$_2$ versus excitation ratios N(4)/N(2)
and N(5)/N(3),
for LMC, SMC, and Galactic data points.  The region shaded in both black
and red includes all points on this plot which are covered by single cloud
models. The region shaded only in red indicates the
area covered by concatenations of any two clouds from
the model grids.  Such concatenations may explain LMC/SMC points that
cannot be reached by single-cloud models.  See \S 6 for discussion.}
\end{figure}


\end{document}